\documentclass[twocolumn,showpacs,preprintnumbers,amsmath,amssymb]{revtex4}
\usepackage{graphicx}% Include figure files
\usepackage{dcolumn}% Align table columns on decimal point
\usepackage{bm}% bold math

\begin{document}

\preprint{APS/xxx}

\title{Complexities of Human Promoter Sequences\\}% Force line breaks with \\
\author{Fangcui Zhao$^1$}
\email{yangzhaon@eyou.com}
\author{Huijie Yang$^{2,3}$}
\email{huijieyangn@eyou.com} \altaffiliation {Corresponding
author}
\author{Binghong Wang$^4$}
\address{$^1$ College of Life Science and Bioengineering,
         Beijing University of Technology,
         Beijing 100022,
         China\\
         $^2$ Department of Physics, National University of
         Singapore, Science Drive 2, Singapore 117543\\
         $^3$ School of Management, University of Shanghai for Science and
         Technology, and Shanghai Institute for Systematic Science, Shanghai 200093, China\\
         $^4$ Department of Modern Physics,
         University of Science and Technology of China,
         Anhui Hefei 230026,
         China
          }
\date{\today}% It is always \today, today,

\begin{abstract}
 By means of the diffusion entropy approach, we detect the scale-invariance
characteristics embedded in the 4737 human promoter sequences. The
exponent for the scale-invariance is in a wide range of $\left[
{0.3,0.9} \right]$, which centered at $\delta _c = 0.66$. The
distribution of the exponent can be separated into left and right
branches with respect to the maximum. The left and right branches
are asymmetric and can be fitted exactly with Gaussian form with
different widths, respectively.

\end{abstract}


\maketitle
\section{Introduction}

Understanding gene regulation is one of the most exciting topics
in molecular genetics \cite{1}. Promoter sequences are crucial in
gene regulation. The analysis of these regions is the first step
towards complex models of regulatory networks.

A promoter is a combination of different regions with different
functions \cite{2,3,4,5}. Surrounding the transcription start site
is the minimal sequence for initiating transcription, called core
promoter. It interacts with RNA polymerase II and basal
transcription factors. Few hundred base pairs upstream of the core
promoter are the gene-specific regulatory elements, which are
recognized by transcription factors to determine the efficiency
and specificity of promoter activity. Far distant from the
transcription start site there are enhancers and distal promoter
elements which can considerably affect the rate of transcription.
Multiple binding sites contribute to the functioning of a
promoter, with their position and context of occurrence playing an
important role. Large-scale studies show that repeats participate
in the regulation of numerous human and mouse genes \cite{6}.
Hence, the promoter's biological function is a cooperative process
of different regions such as the core promoter, the gene-specific
regulatory elements, the enhancers/silencers, the insulators, the
CpG islands and so forth. But how they cooperate with each other
is still a problem to be investigated carefully.

The structures of DNA sequences determine their biological
functions \cite{7}. Recent years witness an avalanche of finding
nontrivial structure characteristics embedded in DNA sequences.
Detailed works show that the non-coding sequences carry long-range
correlations \cite{8,9,10}. The size distributions of coding
sequences and non-coding sequences obey Gaussian or exponential
and power-law \cite{11,12}, respectively. Theoretical model-based
simulations \cite{13,14,15,16} tell us that the parts of the
promoters where the RNA transcription has started are more active
than a random portion of the DNA. By means of the nonlinear
modeling method it is found that along the putative promoter
regions of human sequences there are some segments much more
predictable compared with other segments \cite{17}. All the
evidences suggest that the nontrivial structure characteristics of
a promoter determine its biological functions. The statistical
properties of a promoter may shed light on the cooperative process
of different regions.

Experimental knowledge of the precise 5' ends of cDNAs should
facilitate the identification and characterization of regulatory
sequence elements in proximal promoters \cite{18}. Using the
oligocapping method, Suzuki et al. identify the transcriptional
start sites from cDNA libraries enriched in full-length cDNA
sequences. The identified transcriptional start sites are available
at the Database, http://dbtss.hgc.jp/. \cite{19}. Consequently,
Leonardo et al. have used this data set and aligned the full-length
cDNAs to the human genome, thereby extracting putative promoter
regions (PPRs) \cite{20}. Using the known transcriptional start
sites from over 5700 different human full-length cDNAs, a set of
4737 distinct PPRs are extracted from the human genome. Each PPR
consists nucleotides from $-2000$ to $+1000bp$, relative to the
corresponding transcriptional start site. They have also counted
eight-letter words within the PPRs, using z-scores and other related
statistics to evaluate the over- and under- representations.

In this paper, by means of the concept of diffusion entropy (DE) we
try to detect the scale-invariant characteristics in these putative
promoter regions.

\section{Diffusion Entropy Analysis}

The diffusion entropy (DE) method is firstly designed to capture the
scale-invariance embedded in time series \cite{21,22,23}. To keep
the description as self-contained as possible, we review briefly the
procedures.

We consider a PPR denoted with $Y = \left( {y_1 ,y_2 , \cdots
,y_{3001} } \right)$, where $y_s $ is the element at the position
$s$ and $y_s = A,T,C$ or $G$. Replacing $A,T$ and $C,G$ with $ -
1$ and $ + 1$, respectively, the original PPR is mapped to a time
series $X = \left( {x_1 ,x_2 , \cdots ,x_{3001} } \right)$. There
is not a trend in this series, i.e., $X$ is stationary.

Connecting the starting and the end of $X$, we can obtain a set of
delay-register vectors, which reads,

\begin{equation}
\label{eq1} {\begin{array}{*{20}c}
 {T_1 (t) = \left( {x_1 ,x_2 , \cdots ,x_t } \right)} \hfill \\
 {T_2 (t) = \left( {x_2 ,x_3 , \cdots ,x_{t + 1} } \right)} \hfill \\
 \vdots \hfill \\
 {T_{3001} (t) = \left( {x_{3001} ,x_1 , \cdots ,x_{t - 1} } \right)} \hfill
\\
\end{array} }
\end{equation}

Regarding each vector as a trajectory of a particle in duration of
$t$ time units, all the vectors can be described as a diffusion
process of a system containing $3001$ particles. The initial state
of the system is $\left( {{\begin{array}{*{20}c}
 {T_1 (0)} \hfill \\
 {T_2 (0)} \hfill \\
 \vdots \hfill \\
 {T_{3001} (0)} \hfill \\
\end{array} }} \right) = \left( {{\begin{array}{*{20}c}
 0 \hfill \\
 0 \hfill \\
 \vdots \hfill \\
 0 \hfill \\
\end{array} }} \right)$.

Accordingly, at each time step $t$ we can calculate displacements
of all the particles. The probability distribution function (PDF)
of the displacements can be approximated with $p(m,t)\sim
\raise0.7ex\hbox{${K_m }$} \!\mathord{\left/ {\vphantom {{K_m }
{3001}}}\right.\kern-\nulldelimiterspace}\!\lower0.7ex\hbox{${3001}$}$,
where $m = - t, - t + 1, \cdots ,t$ and $K_m $ is the number of
the particles whose displacements are $m$. It can represent the
state of the system at time $t$.

As a tenet of complexity theory \cite{24,25}, complexity is related
with the concept of scaling invariance. For the constructed
diffusion process, the scaling invariance is defined as,

\begin{equation}
\label{eq2} p(m,t) \approx \frac{K_m }{3001} = \frac{1}{t^\delta
}F\left( {\frac{m}{t^\delta }} \right),
\end{equation}

\noindent where $\delta $ is the scaling exponent and can be
regarded as a quantitative description of the PPR's complexity. If
the elements in the PPR are positioned randomly, the resulting PDF
obeys a Gaussian form and $\delta = 0.5$. Complexity of the PPR is
expected to generate a departure from this ordinary condition,
that is, $\delta \ne 0.5$.

The value of $\delta$ can tell us the pattern characteristics of a
PPR. The departure from the ordinary condition can be described with
a preferential effect. Let the element is $A,T$ (or $C,G$), the
preferential probability for the following element's being $A,T$ (or
$C,G$) is $W_{pre}$. A positive preferential effect, i.e, $W_{pre}
> 0.5$, leads to the value of $\delta$ larger than $0.5$. While a negative
preferential effect, i.e, $W_{pre} < 0.5$, can induce the value of
$\delta$ smaller than $0.5$. Hence, a large value of $\delta$
implies that $A,T$ or $C,G$ accumulate strongly in a
scale-invariance way, respectively.

However, correct evaluation of the scaling exponent is a nontrivial
problem. In literature, variance-based method is used to detect the
scale-invariance. But the obtained Hurst exponent $H$may be
different from the real $\delta $, that is, generally we have$H \ne
\delta $. And for some conditions, the variance is divergent, which
leads the invalidation of the variance-method at all. To overcome
these shortages, the Shannon entropy for the diffusion can be used,
which reads,

\begin{equation}
\label{eq3}
\begin{array}{l}
 S(t) = - \sum\limits_{m = - t}^{t} {p(m,t)\ln p(m,t)} \\
 {\begin{array}{*{20}c}
 \hfill \\
\end{array} } = - \sum\limits_{m = - t}^{t} {\frac{K_m }{3001}} \ln
\left( {\frac{K_m }{3001}} \right). \\
 \end{array}
\end{equation}

This diffusion-based entropy is called diffusion entropy (DE). A
simple computation leads the relation between the scaling
invariance defined in Eq.2 and the DE as,

\begin{equation}
\label{eq4} S(t) = A + \delta \ln t,
\end{equation}

\noindent where $A$ is a constant depends on the PDF. Detailed works
show that DE is a reliable method to search the correct value of
$\delta $, regardless the form of the PDF \cite{26,27,28,29}.

The complexity in the PDF can be catalogued into two levels
\cite{30}, the primary one due to the extension of the probability
to all the possible displacements $m$, and the secondary one due to
the internal structures. Consequently, we should consider also the
corresponding shuffling sequences as comparison.

\begin{figure}
\scalebox{0.7}[0.7]{\includegraphics{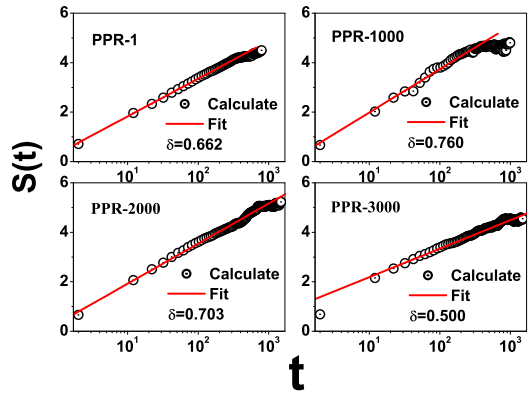}}
\caption{\label{fig:epsart} (Color online) Typical DE results. The
results for the PPRs numbered $1,1000,2000$ and $3000$  are
presented. In considerable wide regions of $t$ , the curves of DE
can be fitted almost exactly with the linear relation in Eq.(4).}
\end{figure}

\begin{figure}
\scalebox{0.7}[0.7]{\includegraphics{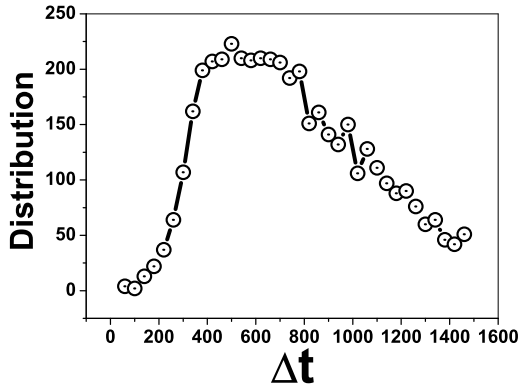}}
\caption{\label{fig:epsart} Distribution of the maximum interval
$\Delta t$  in which one can find scale-invariant characteristics.
Keeping the standard deviation of the fitting result in the range
of $ \le 0.05$ , we can find the maximum intervals  $\Delta t$ for
all the PPRs. The distribution tells us that generally the
scale-invariance can be found over two to three decades of the
scale $t$ . }
\end{figure}

\begin{figure}
\scalebox{0.7}[0.7]{\includegraphics{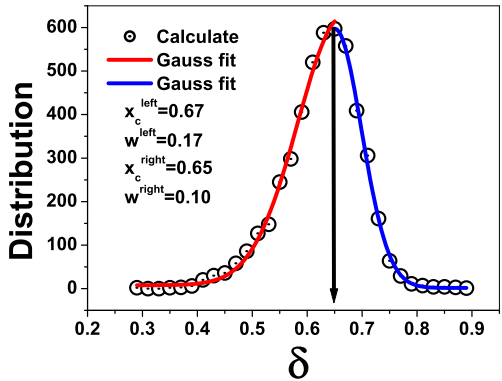}}
\caption{\label{fig:epsart} (Color online) The complex index
$\delta $ distributes in a wide range of $\left[ {0.3,0.9}
\right]$. The distribution can be separated into two branches with
respect to the center $\delta _c = 0.66$. The two branches are
asymmetric and obey exactly the Gauusian function, respectively.
The widths and centers of the left and right branches are
$(w^{left},x_c^{left} ) = (0.17,0.67)$, $(w^{right},x_c^{right} )
= (0.10,0.65)$. The centers coincide with each other, $w^{left}
\approx w^{right} \approx \delta _c = 0.66$. The right branch
distributes in a significant narrow region.}
\end{figure}

\section{Results and Discussions}

The DEs for all the 4737 PPRs are calculated. As a typical
example, Fig.1 presents the DE results for the PPRs numbered
$1,1000,2000$ and $3000$. In considerable wide regions of $t$, the
curves of DE can be fitted almost exactly with the linear relation
in Eq.4.

For each PPR, there exists an interval, $t_0 \sim t_0 + \Delta t$,
in which the PDF behaves scale-invariance. Keeping simultaneously
the standard deviation and the error of the scaling exponent for the
fitting result in the range of $ \le 0.05$ and $\le 0.03$, we can
find the maximum intervals $\Delta t$ for all the PPRs. In the
fitting procedure, the confidence level is set to be $95\%$. The
distribution of $\Delta t$, as shown in Fig.2, tells us that
generally the scale-invariance can be found over two to three
decades of the scale $t$. The concept of DE is based upon
statistical theory, that is, $t_0$ should be large enough so that
the statistical assumptions are valid. To cite an example, we
consider a random series, whose elements obey a homogenous
distribution in $[0,1]$. Only the length of the delay-register
vectors, $t$, in Eq.(1) is large enough, the corresponding PDF for
the displacements, i.e, the summation value of each delay-register
vector, approaches the Gaussian distribution. Consequently, $t_0$ is
not a valuable parameter. The values of $t_0$ for different PPRs are
not presented.

 The resulting scaling exponent $\delta \pm 0.03$
distributes in a wide range of $\left[ {0.3,0.9} \right]$. The
distribution can be separated into two branches with respect to the
center $\delta _c = 0.66$. The two branches are asymmetric and can
be fitted exactly with the Gauusian function, respectively. The
widths and centers of the left and right branches are
$(w^{left},x_c^{left} ) = (0.17,0.67)$, $(w^{right},x_c^{right} ) =
(0.10,0.65)$. That is to say, the centers coincide with each other,
$w^{left} \approx w^{right} \approx \delta _c = 0.66$.
Comparatively, the right branch distributes in a significant narrow
region.

The PPRs are shuffled also. For each PPR, the shuffling result is
obtained by averaging over ten shuffling samples. The scaling
exponents are almost same, i.e., $\delta _{shuffling} = 0.5\pm
0.03$. The detected scale-invariant characteristics are
internal-structure-related.

How to understand the asymmetric characteristic of the distribution
of the complexity index $\delta $ is an interesting problem. In
literature, some statistical characteristics of DNA sequences are
captured with evolution models, such as the long-range correlations
and the over- and under-representation of strings and so on
\cite{31,32,33}. From the perspective of evolution, perhaps the
distribution characteristics may favor a stochastic evolution model.
The initial sequences have same complexity $\delta ^{initial} =
\delta _c = 0.66$. With the evolution processes the sequences
diffuse along two directions, increasing complexity and decreasing
complexity, i.e, the index $\delta $ increases and decreases,
respectively. The diffusion coefficients for the two directions are
significantly different, denoted with $D^{left} \ne D^{right}$.
Based upon the widths of the two branches we can estimate that,
$\raise0.7ex\hbox{${D^{left}}$} \!\mathord{\left/ {\vphantom
{{D^{left}}
{D^{right}}}}\right.\kern-\nulldelimiterspace}\!\lower0.7ex\hbox{${D^{right}}$}
= \raise0.7ex\hbox{${\delta ^{left}}$} \!\mathord{\left/ {\vphantom
{{\delta ^{left}} {\delta
^{right}}}}\right.\kern-\nulldelimiterspace}\!\lower0.7ex\hbox{${\delta
^{right}}$} = 1.7$. It should be noted that, the complexity is
regarded as the departure from the ordinary condition, $\delta=0.5$.
In the totally $4737$ values of $\delta$, only a small portion of
them are less than $0.5$. Accordingly, the PPRs may be catalogued
into two classes, the PPRs with high complexity and the PPRs with
low complexity. The former class evolves averagely with a slow speed
while the later one with a high speed.

In summary, by means of the DE method, we calculate the
complexities of the 4737 PPRs. The distribution of the complexity
index includes two asymmetric branches, which obey Gaussian form
with different widths, respectively. A stochastic evolution model
may provide us a comprehensive understand of these
characteristics.

\section{Acknowledgements}

This work is funded by the National Natural Science Foundation of
China under Grant Nos. 70571074, 10635040 and 70471033, by the
National Basic Research Program of China (973 Program) under grant
No.2006CB705500), by the President Funding of Chinese Academy of
Science, and by the Specialized Research Fund for the Doctoral
Program of Higher Education of China. One of the authors (H. Yang)
would like to thank Prof. Y. Zhuo for stimulating discussions.

\end{document}